\documentclass[compactaffiliation]{Interspeech}

\usepackage{soul}

\interspeechcameraready 

\title{\texorpdfstring{ProMode: A Speech Prosody Model Conditioned on Acoustic and \\Textual Inputs}{ProMode: A Speech Prosody Model Conditioned on Acoustic and Textual Inputs}}

\author[affiliation={1}]{Eray}{Eren}
\author[affiliation={2}]{Qingju}{Liu}
\author[affiliation={3}]{Hyeongwoo}{Kim}
\author[affiliation={2}]{Pablo}{Garrido}
\author[affiliation={1}]{Abeer}{Alwan}

\affiliation{Dept. of Electrical and Computer Engineering}{University of California, Los Angeles}{USA}
\affiliation{}{Flawless AI}{USA}
\affiliation{}{Imperial College London}{UK}
\email{erayeren@g.ucla.edu, qingju.liu@flawlessai.com, hyeongwoo.kim@imperial.ac.uk, pablo.garrido@flawlessai.com, alwan@ee.ucla.edu}

\keywords{speech prosody, speech synthesis, pitch, energy, Perceiver IO}

\begin{document}

\maketitle

\setlength{\parskip}{0pt}
\setlength{\textfloatsep}{4pt plus 1pt minus 1pt}
\setlength{\floatsep}{2.5pt}      
\captionsetup{skip=1pt} 

\begin{abstract}
Prosody conveys rich emotional and semantic information of the speech signal as well as individual idiosyncrasies. 
We propose a stand-alone model that maps text-to-prosodic features such as F0 and energy and can be used in downstream tasks such as TTS. The ProMode encoder takes as input acoustic features and time-aligned textual content, both are partially masked, and obtains a fixed-length latent prosodic embedding. The decoder predicts acoustics in the masked region using both the encoded prosody input and unmasked textual content. Trained on the GigaSpeech dataset, we compare our method with state-of-the-art style encoders. For F0 and energy predictions, we show consistent improvements for our model at different levels of granularity. We also integrate these predicted prosodic features into a TTS system and conduct perceptual tests, which show higher prosody preference compared to the baselines, demonstrating the model’s potential in tasks where prosody modeling is important. 
\end{abstract}

\section{Introduction}
Speech prosody is manifested by changes in pitch (fundamental frequency), energy, and phoneme or word durations. These changes in pitch/energy/duration can help convey the meaning of a sentence, and emotion of the speaker. Hence, modeling prosody accurately can, for example, make synthetic speech sound more expressive. 

Speech prosody modeling is often embedded into other related tasks, such as learning emotional representations in speech emotion recognition (SER) \cite{chen2023exploringwav2vec20finetuning, baevski2020wav2vec20frameworkselfsupervised} and explicit prosodic feature prediction in speech synthesis (TTS) \cite{ren2022fastspeech2fasthighquality, jiang2023fluentspeechstutterorientedautomaticspeech, li2023styletts2humanleveltexttospeech}. For example, the method in \cite{chen2023exploringwav2vec20finetuning} finetunes a pretrained self-supervised learning (SSL) Wav2Vec 2.0 \cite{baevski2020wav2vec20frameworkselfsupervised} model in the downstream SER task. On the other hand, Emotion2vec \cite{ma2023emotion2vecselfsupervisedpretrainingspeech} adopts a self-supervised online distillation strategy on unlabeled emotion data before the downstream SER, rather than using mainstream SSL models \cite{baevski2020wav2vec20frameworkselfsupervised, hsu2021hubertselfsupervisedspeechrepresentation, Chen_2022} pretrained on LibriSpeech-like datasets.
Both approaches take as input raw audio signals. Some of the latest TTS methods involve explicit pitch and energy modeling \cite{ren2022fastspeech2fasthighquality, jiang2023fluentspeechstutterorientedautomaticspeech, li2023styletts2humanleveltexttospeech}, where these prosodic features are often directly conditioned on latent embeddings extracted from Mel-spectrograms and aligned text. The method in \cite{Chen_2022} generates style-agnostic prosodic variations via a pitch predictor, while \cite{li2023styletts2humanleveltexttospeech} proposes a generalized speech style, representing speech characteristics beyond the phonetic content, which can predict pitch and energy associated with a text sequence. The method in \cite{jiang2023fluentspeechstutterorientedautomaticspeech} estimates masked prosodic features by implicitly inferring information from the unmasked context. A similar learning strategy is adopted in the pitch/duration predictor in \cite{shen2023naturalspeech2latentdiffusion}, where a prompt speech encoder implicitly models the prosodic information. Instead of predicting prosodic features of pitch or energy, MegaTTS \cite{jiang2023megattszeroshottexttospeechscale} applies an autoregressive (AR) prosody large language model (LLM) to the quantized prosody codes aggregated at the phoneme level.

The aforementioned (sub-)prosody models do facilitate their related tasks, either SER or TTS; however, they are formulated differently, and are not task-agnostic prosody models. An exception is \cite{hu2023prosodybert}, which is a prosody model that also targets TTS applications. The authors proposed to make use of several speech prosody inputs such as the normalized cross-correlation function, F0, energy, and Mel-spectrogram below 500 Hz. After processing the speech waveform, they utilized a BERT encoder to process these prosodic inputs to generate discrete prosody codes. They used these prosody codes to train and improve an existing TTS system. However, their prosody model lacks textual information, which can also contain prosodic information. Their method also requires training a TTS system to produce prosody-aware speech from text. 

To address these limitations, we propose a zero-shot 
(for both prosody and speaker) and stand-alone speech prosody model utilizing the Perceiver IO structure \cite{jaegle2022perceiveriogeneralarchitecture}, hereafter referred to as ProMode. 
The model's input consists of masked acoustic and textual features, and the model encodes them into latent prosody embeddings via a Perceiver-based encoder. These prosody embeddings reconstruct prosody of the masked region given contexts in the unmasked regions, with two decoders, one conditional and one unconditional to the unmasked text input, aiming for better acoustic input utilization.  
In addition, we modified AdaLN-zero \cite{chen24f5tts} in the conditional decoder to be temporally dependent. 

\begin{figure*}[t]
  \centering
  \includegraphics[width=0.82\textwidth]{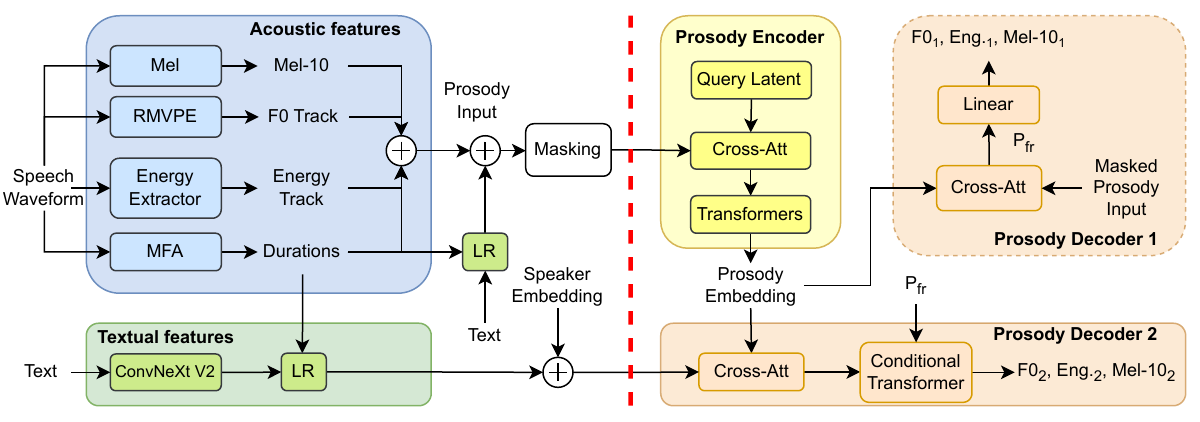}
  \caption{\fontsize{8.5pt}{10pt}\selectfont{The proposed ProMode features the acoustic and textual feature extraction (left to the red dashed line) and the Perceiver IO architecture (right). Different features are projected to embeddings and stacked together (plus-circle). The modified Perceiver has two decoders, with and without textual conditions. Note, Decoder 1 is an axillary decoder to boost convergence of Decoder 2, and only the latter will be kept in the inference stage. Speaker embeddings are obtained from ECAPA2 model~\cite{thienpondt2024ecapa2hybridneuralnetwork}}.
  }
  \label{fig:overall}
\end{figure*}

\section{Methods}
\label{section:met}
\subsection{Overview}
\label{section:met_overview}
ProMode is a zero-shot (for both prosody and speaker) and stand-alone prosody model, which takes a reference input speech and text, and predicts the prosody for a new text and speaker. This enables prosody continuation for a new text given a reference speech signal without requiring training with a TTS model. 
Its topology is illustrated in Fig.~\ref{fig:overall} consisting of: feature preprocessor, Perceiver-based prosody encoder and decoders.

\subsection{Feature Preprocessing}
\label{section:met_preproc}
Speech waveforms were 8 kHz in bandwidth. We extract the Mel-spectrogram, pitch (F0), and energy from the speech signal on a frame-by-frame basis. For noise robustness, we used a noise-robust RMVPE pitch extraction algorithm \cite{Wei_2023}.

Before calculating the energy, the waveform is denoised~\cite{denoiser} (no denoising for F0 extraction since it can distort F0 values). Then, the energy values were obtained in the frequency domain using the short-time-fourier-transform. 

We set energy in non-speech regions to 0 with a noise-robust voice activity detector~\cite{silero}. Finally, we smoothed the logarithmic ($\log_2$) energy values using Savitsky-Golay interpolation~\cite{savitzky1964smoothing}. Since prosodic information is mainly manifested in the low-frequency regions, we calculated the Mel-spectrogram only from the first 10 Mel filters, out of 80 Mel filters; that is 10-dimensional (10D) spanning the low frequencies (less than $380$ Hz).

Phoneme durations contain information related to speech tempo, and also alignment information between frame-level (fr.-level) acoustics and phoneme-level (ph.-level) content, thus they are also extracted, via Montreal forced aligner (MFA) \cite{mcauliffe17_interspeech}. The above features are projected to the embedding space spanning the same dimension and concatenated together. The transcript text is first phonemised via a grapheme to phoneme (G2P) model, and length-regulated (LR), i.e. repeated depending on the phoneme duration, to align with the acoustic features. The aligned text embeddings are further concatenated with the audio acoustics as input to the prosody encoder. 

\subsection{Prosody Encoder}
\label{section:met_prosenv}
To deal with audio sequences of varying length, the standard Perceiver IO encoder \cite{jaegle2022perceiveriogeneralarchitecture} is employed, composed of a cross attention layer followed by stacked transformer layers. The dimension and length of the output prosody embedding depend on the size of the learnable query latent. We randomly mask 60\% of the input audio frames in a way that the boundary of each masked block overlapped with a phoneme boundary, i.e. a phoneme is either fully masked or fully available.

\subsection{Prosody Decoder with Acoustic Only Loss}
\label{section:met_loss}
We modify the standard prosody decoder to adapt to our prosody feature prediction tasks. To generalize to masked/new content, we use the textual inputs as queries to attend the latent embeddings. To exploit long temporal relation in the text sequence, we employ stacked convolutional layers ConvNeXt V2 \cite{convnext} to the text conditions before applying cross-attention. 
Our preliminary experimental findings revealed that acoustic information conditioning of the decoder tends to collapse during training. That is, the prosody embeddings can be ignored considerably; and hence, the decoder focuses only on the textual feature input of the decoder. 

To prevent this collapse and boost mode coverage, we propose to use two decoders, one conditional and one unconditional on the text input. This is also inspired by the classifier free guidance in diffusion models \cite{ho2022classifierfreediffusionguidance}, where the generations mix the score estimate of a conditional model and an unconditional model, e.g. for image generation \cite{ho2022classifierfreediffusionguidance} and audio generation  \cite{le2023voiceboxtextguidedmultilingualuniversal}. We keep both conditional (Prosody Decoder 2, PD2) and unconditional (Prosody Decoder 1, PD1) predictions, where decoder PD1 omits the unmasked text condition. For both PD1 and PD2, we calculate losses between the ground-truth and the associated predictions, where losses from PD1 serve as auxiliary ``acoustic only loss (AOL)''. Although the prosody encoder processes the context (regions other than masked region) textual information as well, the inputs to the encoder are dominated by acoustical features. Therefore, the PD1 module focuses mostly on the acoustic information with the help of the AOL. To obtain the outputs of the PD1 module, the output ($P_{fr}$ vectors) of the cross-attention between the masked prosody inputs and prosody embeddings are projected with linear layers. At inference time, we drop the first unconditional decoder PD1 and use results predicted via the conditional decoder PD2. 

\subsubsection{Modified adaLN-zero}
\label{section:met_modadaln}
PD2 employs a conditional transformer, resembling the adaLN-zero conditioning in F5-TTS \cite{chen24f5tts}. 
In \cite{chen24f5tts}, the authors proposed to learn the scale/shift parameters before and after the multi-head self-attention (MHSA) module and feed-forward module inside the transformer layer. This results in 6 scale/shift parameters inside the transformer layer. However, these scale/shift parameters are global, not offering the flexibility to learn temporally dependent scale/shifts based on the given condition, which is prosody embeddings in our case. Therefore, we modify adaLN-zero to predict the 6 scale/shift parameters for all time steps by adding an additional cross-attention module that attends to prosody embedding (key) and masked prosody input (query) to predict $T\times6$ scale/shift parameters. This cross-attention maps the fixed-length latent temporal dimension of prosody embeddings into length $T$. 

\vspace{-0.12in}
\section{Experimental Settings}
\label{section:exp}
\subsection{Data and Setup}
We run our experiments on GigaSpeech~\cite{chen2021gigaspeechevolvingmultidomainasr}, which contains English spoken in read and spontaneous styles. We trained ProMode on the M subset of GigaSpeech, ($\sim1000$ hours). We filtered out samples if their length is not between 2 and 22 seconds, leaving $\sim750$ hours. The Dev/Test split in GigaSpeech (12/40 hours) was used as validation/test data.

Audio features were extracted at 11.6 ms per frame. We used the first 10D from the Mel-spectrogram. We trained our own aligner models on the same training set with MFA. 

We incorporated voiced or unvoiced (vuv) feature as a binary input (also as output) to the extracted acoustic features. These extracted features were projected onto an embedding space spanning 128D, via stacked Linear Layer, ReLU and LayerNorm. After stacking the above features, they were further projected onto 768D. Then, they were convolved with 4 ConvNeXt V2 layers \cite{convnext}, each with 512D hidden dimension and kernel size of 7. We used a similar setting for textual features' ConvNeXt V2, and  rotary positional encodings were incorporated before the convolutions. The resulting vector was fed into the prosody encoder as prosody input. We used Perceiver IO base parameter in \cite{jaegle2022perceiveriogeneralarchitecture} for our prosody encoder; however, we used 8 attention heads and 18 self-attention layers. Speaker embeddings were extracted via the ECAPA2 model \cite{thienpondt2024ecapa2hybridneuralnetwork}. For each output of the PD2 module, we used 4 layer and 512D hidden dimensonal conditional transformer blocks \cite{chen24f5tts} with 4 attention heads (each head 64D). Finally, after predictions, we combined the PD1 and PD2 losses. That is, L1 losses for F0s, MSE losses for Mel-10 and energy, and binary cross-entropy losses for vuv. 

We compare ProMode with  three counterpart prosody encoders: StyleTTS2 encoder \cite{li2023styletts2humanleveltexttospeech}, Wav2Vec2-SER \cite{chen2023exploringwav2vec20finetuning}, and Emotion2Vec \cite{ma2023emotion2vecselfsupervisedpretrainingspeech}. For fair comparisons,  we combine these baselines with ProMode decoders, and train on the same training set with the same loss. The pretrained baseline models are frozen during training. The  baselines with re-trained decoders are denoted as StyleTTS2*, Wav2Vec2-SER* and Emotion2Vec*. We trained ProMode for 250k iterations with batch size of 32 using Adam optimizer~\cite{adam}. For the baselines, since the validation metrics became inferior with the same iterations, we applied early stopping around 60k to obtain better models.

\subsection{Evaluation Metrics}
To assess the predicted F0 and energy values, several objective metrics were used. For both F0 and energy, the evaluations were performed at the fr.-level and ph.-level. To ensure alignment between ground truth and predicted values, ground truth durations were input to ProMode and the baselines. All objective metrics were calculated after dynamic time warping (DTW) of the predicted values with respect to ground truth values. 

Fr.-level metrics consist of root mean square error (RMSE) \cite{ li2023styletts2humanleveltexttospeech, shen2023naturalspeech2latentdiffusion, chen22unsuper, lei22towards} and mean absolute error (MAE) \cite{chien21hierarc} of the F0 and energy values. Raw pitch accuracy (RPA) and raw chroma accuracy (RCA) metrics~\cite{mireval} were calculated for F0 similar to the way they are used for analyses of speech, singing voices, or music~\cite{kum2019joint, crep}. In~\cite{kum2019joint, crep}, RPA and RCA are obtained from ground truth F0 compared to predicted F0 from the given audio; however, we predict F0 from a new text, which is more challenging (hence, low values are expected). For energy evaluations, we used the logarithmic ($\log_2$) scale mean absolute error (MAE$_{\log}$) \cite{logeng}. The ph.-level metrics include MAE, as well as the difference of ph.-level statistics (means or $\mu$, and variances or $\sigma$) between the ground truth and predicted values of F0 and energy, similar to \cite{shen2023naturalspeech2latentdiffusion}. For F0, we used the ground truth voiced frames for calculating every metric, except RMSE.

\setlength{\tabcolsep}{2pt} 
\begin{table*}[h!] 
   \caption{\fontsize{8.2pt}{10pt}\selectfont{F0 raw pitch accuracy (RPA,\%), raw chroma accuracy (RCA,\%), root mean square error, fr.-level/ph.-level mean absolute error (MAE), fr.-level difference of means ($\mu$), fr.-level difference of standard deviations ($\sigma$), ph.-level difference of means difference ($\mu$), ph.-level difference of standard deviations ($\sigma$), $log_2$ scale mean absolute error (MAE$_{\log}$). Asterisk (*) means that we further trained the baseline models' embeddings on our data, our decoder, and our objective. We underline statistically insignificant values of ProMode ($p-value>0.05$ and with respect to the best baseline). Bold values indicate the best performance.}}
  \label{tab:performance_metrics}
  \centering
  \footnotesize
  \begin{tabular}{lcccc|ccc|cc|ccc|ccc|cc}
    \toprule
    \textbf{Model} & \multicolumn{9}{c|}{\textbf{F0}} & \multicolumn{8}{c}{\textbf{Energy}} \\
    
    \cmidrule(lr){2-10} \cmidrule(lr){11-18} & 
    \multicolumn{4}{c|}{\textbf{Fr.-Level}} & 
    \multicolumn{3}{c|}{\textbf{Ph.-Level}} & 
    \multicolumn{2}{c|}{\textbf{Fr.-Level}} & 
    \multicolumn{3}{c|}{\textbf{Fr.-Level}} & 
    \multicolumn{3}{c|}{\textbf{Ph.-Level}} & 
    \multicolumn{2}{c}{\textbf{Fr.-Level}} \\
    
     \cmidrule(lr){2-6} 
     \cmidrule(lr){7-8} 
     \cmidrule(lr){9-10} 
     \cmidrule(lr){11-14} 
     \cmidrule(lr){15-16} 
     \cmidrule(lr){17-18}
     
    & \textbf{RPA}$\uparrow$ & \textbf{RCA}$\uparrow$ & \textbf{RMSE}$\downarrow$ & \textbf{MAE}$\downarrow$ & \textbf{MAE}$\downarrow$ & $\bm{\mu}$$\downarrow$  & $\bm{\sigma}$$\downarrow$ & $\bm{\mu}$$\downarrow$ & $\bm{\sigma}$$\downarrow$ & 
    
    \textbf{MAE$_{\log}$}$\downarrow$ & \textbf{RMSE}$\downarrow$ & \textbf{MAE}$\downarrow$ & \textbf{MAE}$\downarrow$ & $\bm{\mu}$$\downarrow$  & $\bm{\sigma}$$\downarrow$ & $\bm{\mu}$$\downarrow$ & $\bm{\sigma}$$\downarrow$ \\
    
    \midrule
    
    StyleTTS2* & 40.2 & 39.8 & 22.3 & 15.7 & 14.9 & 10.7 & 10.8 & 10.2 & 10.9
    & 0.681 & 3.15 & 1.27 & 1.24 & 1.14 & 1.82 & 0.98 & 2.14 \\

    Emotion2Vec* & 35.1 & 35.0 & 25.1 & 19.0 & 18.8 & 12.9 & 13.9 & 12.5 & 14.2
    & 0.697 & 3.32 & 1.36 & 1.35 & 1.15 & 1.88 & 1.07 & 2.25  \\

    Wav2Vec2-SER* & 39.9 & 39.6 & 22.0 & 16.3 & 16.0 & 9.12 & 13.1 & 8.60 & 12.8
    & 0.654 & 3.27 & 1.32 & 1.29 & 1.06 & 1.86 & 0.99 & 2.20  \\

    ProMode (Ours) & \textbf{43.9} & \textbf{43.6} & \textbf{21.5} & \textbf{14.9} & \textbf{14.7} & \textbf{8.23} & \textbf{9.94} & \textbf{7.74} & \textbf{9.89}
    & \textbf{0.633} & \textbf{2.97} & \textbf{1.18} & \textbf{\underline{1.16}} & \textbf{\underline{0.96}} & \textbf{1.66} & \textbf{0.87} & \textbf{1.96} \\

    \hline
    \multicolumn{18}{l}{\quad \textbf{Ablations: ProMode -(input/module taken out)}} \\
    \hline
    \quad -F0 & 42.3 & 42.0 & 22.2 & 15.8 & 15.6 & 10.4 & 12.4 & 9.86 & 12.5
    & 0.658 & 3.20 & 1.29 & 1.29 & 1.12 & 1.85 & 1.02 & 2.18  \\

    \quad -Eng & 41.2 & 40.9 & 21.9 & 16.0 & 15.6 & 9.39 & 11.7 & 8.98 & 11.7
    & 0.698 & 3.13 & 1.27 & 1.24 & 1.05 & 1.8 & 0.98 & 2.09  \\

    \quad -Dur & 40.2 & 39.9 & 22.8 & 16.8 & 16.6 & 11.0 & 12.5 & 10.5 & 12.6
    & 0.628 & 3.25 & 1.32 & 1.31 & 1.13 & 1.85 & 1.05 & 2.20  \\

    \quad -Context Text & 41.9 & 41.7 & 21.6 & 15.5 & 15.1 & 8.70 & 10.7 & 8.28 & 10.7
    & 0.593 & 2.69 & 1.05 & 1.00 & 0.75 & 1.36 & 0.70 & 1.64 \\

    \quad -Mel10 & 42.2 & 41.9 & 21.9 & 15.6 & 15.3 & 9.17 & 11.3  & 8.70 & 11.2
    & 0.643 & 2.75  & 1.09 & 1.05 & 0.83 & 1.40 & 0.75 & 1.66 \\

    \quad -AOL & 25.8 & 26.0 & 30.1 & 25.3 & 25.3 & 23.2 & 12.9 & 22.7 & 13.5
    & 0.787 & 3.14 & 1.27 & 1.25 & 1.06 & 1.72 & 0.98 & 2.06 \\

    \quad -M. adaLN-zero & 40.9 & 40.7 & 22.0 & 15.6 & 15.4 & 8.67 & 11.2 & 8.22 & 11.3
    & 0.709 & 3.13 & 1.26 & 1.24 & 1.05 & 1.72 & 0.97 & 2.07 \\

    \bottomrule
  \end{tabular}
\end{table*}

\subsection{Downstream TTS}
\label{sec:downstreamtts}
ProMode is a stand-alone model to predict prosodic features, offering the opportunity to support downstream tasks such as TTS that require these features. We integrate ProMode into FluentSpeech \cite{jiang2023fluentspeechstutterorientedautomaticspeech}, a TTS method featuring a diffusion approach \cite{ho2020denoisingdiffusionprobabilisticmodels}. We replace its build-in pitch predictor with ProMode-predicted pitch. When testing on GigaSpeech-Test, we divide a sentence into two parts. The first half is used as an audio prompt to obtain the prosody embedding via ProMode Encoder. The transcript associated with the second half is used as a textual condition for ProMode Decoder, and the predicted pitch will be used in the TTS system. The same procedure is also applied to StyleTTS2*, Wav2Vec2-SER* and Emotion2Vec* for comparisons. To avoid the selected TTS becoming a bottleneck, we upscaled and retrained FluentSpeech  on a larger set (GigaSpeech-XL). FluentSpeech relies on an independent vocoder to convert Mel spectrograms to time waveforms, and we selected the pretrained BigVGAN \cite{lee2023bigvganuniversalneuralvocoder} vocoder that is band-limited to 8 kHz. To evaluate the performance of the TTS system that integrates ProMode or other baselines, we use UTMOS \cite{saeki2022utmosutokyosarulabvoicemoschallenge} to evaluate overall audio quality, cosine speaker similarity SECS~\cite{xttsv2}, WER($\%$) with Whisper-large for speech intelligibility, and AutoPCP \cite{seamless} for prosody similarity. Samples are available at \textcolor{blue}{\href{https://promode8272.github.io/promode/index.html}{https://promode8272.github.io/promode/index.html}}.

\section{Results and Discussion}
\label{section:results}
\subsection{Objective Results of F0 and Energy Prediction}
Objective evaluations of F0 and energy predictions are shown in Table~\ref{tab:performance_metrics}. Our model shows consistent advantages over the baselines, in both pitch and energy accuracy at different levels. Since the same decoder is employed for all baselines, the performance improvement is introduced by our prosody encoder. Of the three baselines with retrained decoders, the TTS-based method StyleTTS2* produced results that were relatively closer to ProMode, while the two SER-based methods performed much worse. Interestingly, although Emotion2Vec* was finetuned with more labeled data than Wav2Vec2-SER*, it exhibited lower performance.

\subsection{Downstream TTS}
\subsubsection{Objective Results of Downstream TTS}
When integrating ProMode and baselines into the same downstream TTS task, we evaluate the synthesized speech in intelligibility, quality, and speaker/prosody similarity, as introduced in Sec.~\ref{sec:downstreamtts}. The average scores are listed in Table \ref{tab:tts}. 

It can be observed that while all methods  effectively preserve speaker identity with higher SECS, ProMode yields more intelligible speech with a lower WER, and higher naturalness as reflected by larger UTMOS scores. Moreover, ProMode is better at capturing the overall style and prosody of the prompt audio. These findings are consistent the results in Table \ref{tab:performance_metrics}. We directly fed the ground truth Mel-spectrogram to the BigVGAN vocoder, denoted as ``GT (voc.)'', indicating the upper thresholds of the synthesized speech. ProMode and all baselines have achieved comparable or even better WER than ``GT (voc.)''; a similar result was observed in \cite{voicecraft}.

\begin{table}[h]
    \centering
    \footnotesize
    \caption{\fontsize{8.2pt}{10pt}\selectfont{Objective evaluations on a downstream TTS task. ``GT (voc.)'' indicates vocoded ground truth speech. We underline statistically insignificant values of ProMode ($p-value>0.05$ and with respect to the best baseline).}}
    \begin{tabular}{ccccc}
        \hline
        \textbf{Model} & \textbf{SECS$\uparrow$} & \textbf{WER (\%)$\downarrow$} & \textbf{UTMOS$\uparrow$} & \textbf{AutoPCP$\uparrow$} \\ 
        \hline
        GT (voc.)     & -  & 4.52  & 3.13  & -  \\
        \hline
        FluentSpeech    & 0.81  & 4.41  & 3.00  & 2.47  \\ 
        StyleTTS2*      & 0.79  & 4.51  & 2.95  & 2.48  \\ 
        Emotion2Vec*    & 0.80  & 4.25  & 2.73  & 2.33  \\ 
        Wav2Vec2-SER*   & 0.79  & 4.56  & 2.76  & 2.34  \\ 
        ProMode (Ours)  & \textbf{0.83}  & \underline{\textbf{3.99}}  & \textbf{3.10}  & \textbf{2.64}  \\  
        \hline
    \end{tabular}
    \label{tab:tts}
\end{table}
\begin{figure}[t]
    \centering
    \includegraphics[width=0.88\columnwidth]{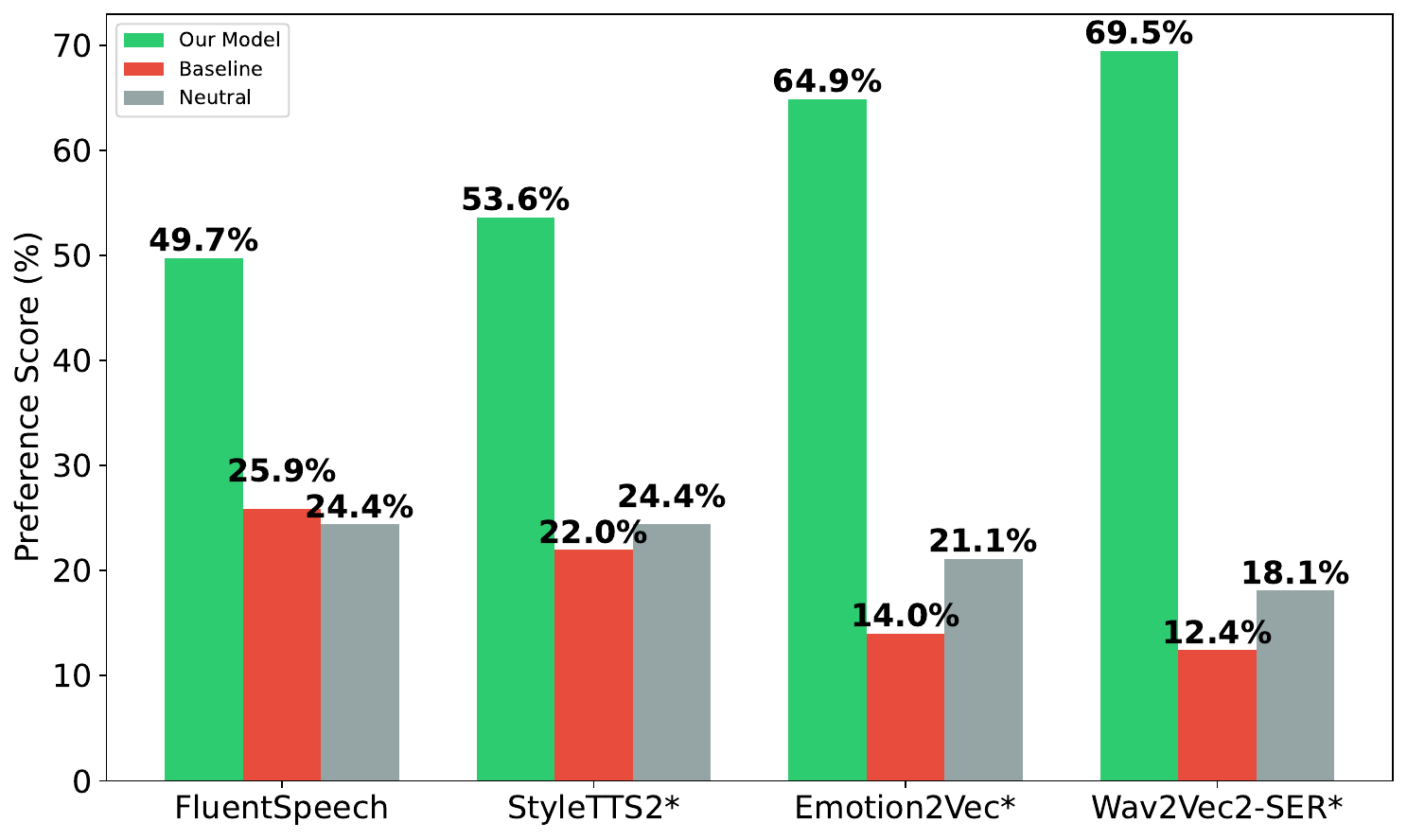}
    \caption{\fontsize{8.2pt}{10pt}\selectfont{ABX preference scores of prosody compared to the baselines. The green/red/gray colors denote ProMode/Baseline/Neutral. Neutral indicates that A and B are equally similar to the ground truth speech. ProMode preferences are statistically significant ($p-value<0.05$) with respect to each baseline.}}
    \label{fig:abx_scores}
\end{figure}

\subsubsection{Subjective Results of Downstream TTS} 
A perceptual ABX listening test was conducted. Participants were instructed to compare a reference signal with two synthetic speech samples (A and B) and select the sample that better preserves prosody. In total 39 native English-speaking listeners were recruited from the USA via Prolific \cite{Prolific2025}. Each participant evaluated 40 test samples. Listening test results are shown in Fig \ref{fig:abx_scores}. Listeners showed a strong preference for ProMode, compared to the three baselines and FluentSpeech. This indicates that predicted pitch values of ProMode lead to significantly better prosody perception of the synthesized speech.

\subsection{Ablations}
To measure the importance of the ProMode inputs on prediction accuracy, we conducted an ablation study as shown in Table~\ref{tab:performance_metrics}. We excluded each component one-by-one, and trained the network without that component. The components are: F0, energy (Eng), duration (dur), text, Mel-spectrogram (Mel10), and acoustic only loss (AOL). Removal of any input feature led to performance degradation to some degree, with AOL being the most critical component.  When AOL is removed, it causes significant degradations in all metrics. The reason is that the absence of AOL leads to not accounting for acoustic inputs (prosody embeddings), and only attending to textual inputs. Also, use of adaLN-zero instead of modified (M.) version leads to inferior results, indicating its effectiveness.

The exclusion of the duration inputs also leads to considerable degradation for both F0 and energy predictions compared to the exclusion of F0 or energy inputs. This occurs when the network predicts F0 and energy from other inputs such as Mel10, but it cannot recover the duration information. Also, exclusion of context text and Mel10 leads to worse F0 predictions, but energy values improve. 

\section{Conclusion}
\label{section:conclusion}
We presented ProMode, a zero-shot, stand-alone speech prosody model built upon the Perceiver IO architecture. ProMode takes masked acoustic and textual features as input and predicts F0 and energy for the masked regions, leveraging contextual information from the unmasked segments. Unlike many existing prosody models that are tightly coupled with specific downstream tasks (like SER or TTS), ProMode is designed to be general-purpose, allowing its predicted F0 and energy to be used in a variety of applications. The proposed dual-decoder approach in the proposed Perceiver-based architecture, combining both conditional and unconditional decoders, along with an auxiliary acoustic-only loss, enhances the model's ability to capture prosodic variations.

Our experiments on GigaSpeech showed ProMode's superior performance compared to several baselines in predicting both F0 and energy. Also, we integrated ProMode into a state-of-the-art TTS system (FluentSpeech) and showed its ability to improve the prosody of synthesized speech, suggesting ProMode's potential impact on downstream applications requiring accurate prosody modeling. Future work will explore extensions to other languages and speech-related tasks such as speech editing. 

\textbf{Acknowledgements:}
We would like to thank Dr. Dana Watson for her valuable editorial assistance.

\bibliographystyle{IEEEtran}
\bibliography{mybib}

\end{document}